\documentclass[]{pasj00}
\SetRunningHead{Y. Ohyama}{IRC Spectroscopy}

\Received{April 6, 2007}
\Accepted{June 21, 2007}

\begin{document}

\title{Near-infrared and Mid-infrared Spectroscopy with the Infrared Camera (IRC) for AKARI}

\author{
        Youichi \textsc{Ohyama},\altaffilmark{1}
        Takashi \textsc{Onaka},\altaffilmark{2}
        Hideo \textsc{Matsuhara},\altaffilmark{1}
        Takehiko \textsc{Wada},\altaffilmark{1}\\
        Woojung \textsc{Kim},\altaffilmark{1}
        Naofumi \textsc{Fujishiro},\altaffilmark{3}\thanks{
        Present Address is Cybernet systems Co. Ltd., Bunkyo-ku, Tokyo 112-0012, Japan}
        Kazunori \textsc{Uemizu},\altaffilmark{1}
        Itsuki \textsc{Sakon},\altaffilmark{2}\\
        Martin \textsc{Cohen},\altaffilmark{4} 
        Miho \textsc{Ishigaki},\altaffilmark{5}\thanks{
        Present Address is Astronomical Institute, Tohoku University, Aoba-ku, Sendai 980-8578, Japan}
        Daisuke \textsc{Ishihara},\altaffilmark{2}
        Yoshifusa \textsc{Ita},\altaffilmark{1}\\
        Hirokazu \textsc{Kataza},\altaffilmark{1}
        Toshio \textsc{Matsumoto},\altaffilmark{1}
        Hiroshi \textsc{Murakami},\altaffilmark{1}
        Shinki \textsc{Oyabu},\altaffilmark{1}\\
        Toshihiko \textsc{Tanab\'e},\altaffilmark{6}
        Toshinobu \textsc{Takagi},\altaffilmark{1}
        Munetaka \textsc{Ueno},\altaffilmark{7}
        Fumio \textsc{Usui},\altaffilmark{1}\\
        Hidenori \textsc{Watarai},\altaffilmark{8}
        Chris P. \textsc{Pearson},\altaffilmark{1,9}
\and        
        Norihide \textsc{Takeyama},\altaffilmark{10}
        Tomoyasu \textsc{Yamamuro},\altaffilmark{10}\thanks{
        Present Address is OptCraft, Hadano, Kanagawa, 259-1331, Japan}
        Yuji \textsc{Ikeda},\altaffilmark{10}\thanks{
Photocoding, Higashi-Hashimoto, Sagamihara, Kanagawa, 229-1104, Japan}
        }

\altaffiltext{1}{Institute of Space and Astronautical Science, \\
Japan Aerospace Exploration Agency, Sagamihara, Kanagawa 229-8510, Japan}
\email{ohyama@ir.isas.jaxa.jp}
\altaffiltext{2}{Department of Astronomy, Graduate School of Science,
The University of Tokyo, \\
Bunkyo-ku, Tokyo 113-0033, Japan}
\altaffiltext{3}{Department of Physics, Graduate School of Science,
The University of Tokyo, \\
Bunkyo-ku, Tokyo 113-0033, Japan}
\altaffiltext{4}{Radio Astronomy Laboratory, 601 Campbell Hall, 
University of California,\\
Berkeley, CA94720, U.S.A.}
\altaffiltext{5}{Department of Physics, Faculty of Science,
Tokyo Institute of Technology, \\
Meguro-ku, Tokyo 15208551, Japan}
\altaffiltext{6}{Institute of Astronomy, Graduate School of Science,
The University of Tokyo, \\
Mitaka, Tokyo 181-0015, Japan}
\altaffiltext{7}{Department of Earth Science and Astronomy, 
Graduate School of Arts and Sciences,\\
The University of Tokyo, 
Meguro-ku, Tokyo 153-8902, Japan}
\altaffiltext{8}{Office of Space Applications, Japan Aerospace Exploration
Agency, \\
Tsukuba, Ibaraki 305-8505, Japan}
\altaffiltext{9}{ISO Data Center, European Space Agency,
Villafranca del Castillo, P.B.Box 50727,\\
28080 Madrid, Spain}
\altaffiltext{10}{Genesia Corporation, Shimo-renjaku,
Mitaka, Tokyo 181-0013, Japan}


\maketitle

\begin{abstract}

The Infrared Camera (IRC) is one of the two instruments on board the {\it AKARI} satellite.
In addition to deep imaging from 1.8--26.5$\mu$m for the pointed observation mode of the {\it AKARI}, it has a spectroscopic capability in its spectral range.
By replacing the imaging filters by transmission-type dispersers on the filter wheels, it provides low-resolution ($\lambda$/$\delta \lambda \sim$ 20--120) spectroscopy with slits or in a wide imaging field-of-view (approximately 10\arcmin$\times$10\arcmin).
The IRC spectroscopic mode is unique in space infrared missions in that it has the capability to perform sensitive wide-field spectroscopic surveys in the near- and mid-infrared wavelength ranges.
This paper describes specifications of the IRC spectrograph and its in-orbit performance.

\end{abstract}

\section{Introduction}

The Infrared Camera (IRC: \cite{onaka07}) is one of the two instruments on board the {\it AKARI} satellite (\cite{murakami07}), and is mainly designed for deep imaging from 1.8--26.5$\mu$m for {\it AKARI}'s pointed observation mode.
In order to obtain rich multi-color information of astronomical objects at near- and mid-infrared (hereafter, NIR and MIR) wavelengths, the IRC is composed of three channels (NIR, MIR-S, and MIR-L), each equipped with a selectable filter wheel to cover a wide wavelength range.
In addition to the multi-color photometry capability, a spectroscopic mode was also introduced in its spectral range to enhance the IRC's capability of analyzing the spectral energy distribution of brighter sources.

By replacing the imaging filters by transmission-type dispersers on the filter wheels, it provides low-resolution ($\lambda$/$\delta \lambda \sim$ 20--120) slit-less spectroscopy in its imaging field-of-view (FOV) (approximately 10\arcmin$\times$10\arcmin) for the first time in space infrared observations.
The IRAS LRS provided also slit-less spectroscopy (\cite{iras}), but it made observations of anything that was located in the wide
entrance apertures of the separate blue and red spectrometers, whether a single object, or
multiple objects, or even bright diffuse emission.
Deconvolving these entangled spectra was often problematic. 
The IRC slit-less spectroscopy, on the other hand, enables multi-object spectroscopy, providing the advantage of efficient observations of multi-objects in a region, although it suffers from high background, particularly in the MIR where the zodiacal emission is dominant.
While a slit spectrograph, such as the Infrared Spectrograph (IRS) on {\it Spitzer} (\cite{irs}), provides better sensitivity
in high background regions, the IRC slit-less spectroscopy allows a serendipitous survey of interesting objects with blank field observations.
Note that the IRC also has small slit areas for all the three channels to allow spectroscopy of diffuse sources.
Compared to the ISOCAM CVF mode on the ISO satellite (\cite{isocam_cvf}) which also provided multi-object spectroscopic capability, 
the IRC achieves higher sensitivity in a short exposure thanks to the large format arrays with low read noise.

Another advantage of IRC spectroscopy, by comparison with the IRS in particular, is that the IRC can perform NIR spectroscopy 
(2--5$\mu$m), where there are interesting lines, such as H$_2$ $\nu$1--0 rotational lines and hydrogen recombination lines, 
and dust bands, such as 3.1$\mu$m water ice and the 3.3 and 3.4$\mu$m unidentified infrared bands.

This paper describes specifications of the IRC spectrograph and its in-orbit performance.
See \citet{onaka07} for details of the general IRC design and its imaging performance.

\section{Design and Characteristics of the IRC Spectroscopic Mode}

The IRC spectroscopic mode shares most of the basic hardware and software designed for its imaging mode (e.g., camera optics, electronics for controlling the cameras, software for controlling/reading out detector arrays, communication with the satellite attitude control system).
Under these hardware/software constraints, we decided to focus mainly on slit-less spectroscopy in its imaging FOV to make the spectroscopic 
mode more attractive and powerful in terms of science, be simple and reliable in terms of instrument development and operation, and not compromise any capabilities and/or performance of the IRC imaging mode.
In order to fully utilize its capability for spectroscopic surveys by minimizing loss of spectral information due to source overlap or contamination, the size of the spectral images along the dispersion direction is restricted to be about one fifth of the FOV.
Therefore, only low spectral resolution is available for slit-less spectroscopy.

One must also acknowledge some weaknesses of slit-less spectroscopy.
One is the higher background in spectroscopic images.
Although the MIR background from space is faint enough to make the slit-less spectroscopy viable, the spectral images suffer higher background noise, because emission from compact sources is dispersed relative to broad band imaging, while that of the background changes very little with respect to broad band imaging.
Another limitation is poorer uniformity of calibration.
Positional dependence of the spectral calibration should be taken into account to obtain uniformly calibrated spectra observed with slit-less spectroscopy, and this 
requires much more detailed calibration information than for slit spectroscopy.
However, since the IRC does not have any good built-in calibration lamps, we must observe celestial objects as our calibrators 
to construct the calibration database.
In reality, because of the limited cryogenic lifetime of {\it AKARI} and very tight target visibility constraints posed by the nature of 
its orbit (solar-synchronous near-earth polar-orbit for the all-sky survey: \cite{murakami07}), the available calibrators are rather limited.
In spite of these difficulties, however, we have so far achieved good calibration performance, as detailed in the following sections,
so that we can indeed conduct deep blank-field spectroscopic surveys across the entire IRC wavelength range (\cite{wada07}).

\section{Instrumentation Specific to the IRC Spectroscopy Mode}

To add the spectroscopic functionality to the IRC, we developed and installed dispersers and focal-plane aperture masks with slit areas.
Here we describe them in more detail.
See also \citet{onaka07} for more information on the IRC in general.

\subsection{Dispersers}

We developed transmissive dispersers of direct-vision type, and mounted them on the filter wheels.
Basic specifications of the dispersers were determined by requiring that the dispersed image length be as small as one fifth of the 
total array size (except for a higher dispersion grism in the NIR channel), while still covering the entire wavelength range of the channels.
The dispersion (wavelength increment per pixel) is simply determined by the required wavelength coverage and the spectral image length, and the spectral resolving power is determined by the point spread function (PSF) size in the imaging mode that is set either by wavefront error (at $\lesssim 7.3\mu$m) or the diffraction limit (at $\gtrsim 7.3\mu$m) of the telescope optics (\cite{kaneda07}).
The dispersers thus designed provide rather low resolving power ($R=\lambda$/$\delta \lambda$) of $\sim$ 20--50.
In addition, a higher-power disperser for the NIR channel was introduced.
Utilizing the full dimension of the NIR detector array, its resolving power was set to be as large as possible, $R \simeq 120$, determined mostly by the array size and the PSF size.

\begin{figure*}
	\begin{center}
	\FigureFile(120mm,50mm){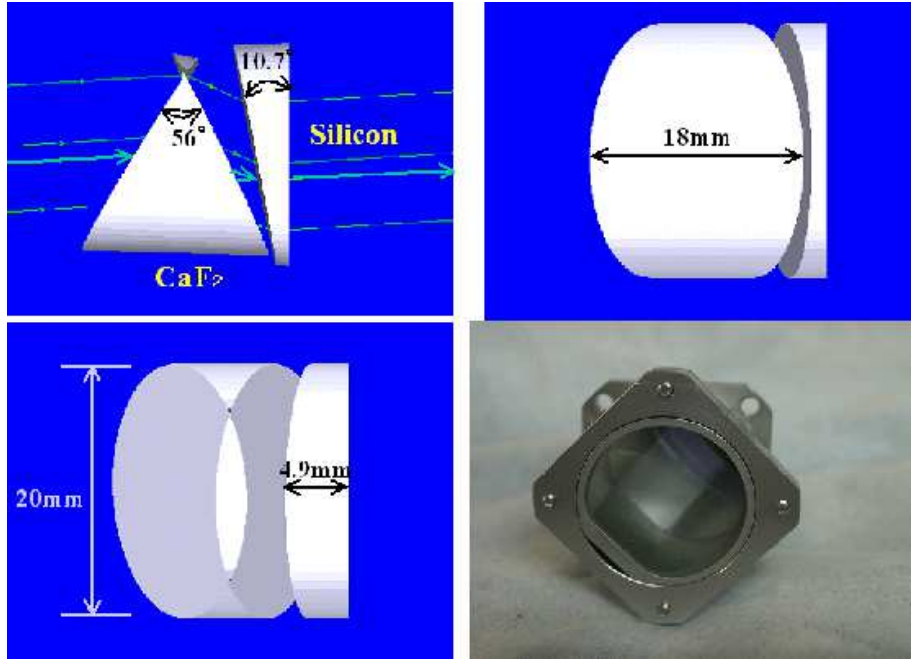}
	\end{center}
\caption{Drawings and a picture of NP.
($upper$ $left$, $upper$ $right$, and $lower$ $left$) Schematic drawings of the prisms comprising the NP.
($lower$ $right$) A picture of the assembled NP in its holder, seen from the ray incident side.
}\label{fig:NP}
\end{figure*}

For low-dispersion spectroscopy, we chose a prism for the NIR channel and grisms for MIR-S and MIR-L channels.
The NIR prism (NP) simultaneously covers the entire wavelength coverage of the channel (1.8--5.5$\mu$m).
The NP is comprised of two individual prisms to form a single low-dispersion direct-vision prism (Figure \ref{fig:NP}).
For the MIR-S channel, since the channel covers a wide spectral range of more than an octave (5--13$\mu$m), two grisms (SG1 and SG2) 
handle the short and long halves of the wavelength coverage, respectively, in the first-order.
Similarly, for the MIR-L channel, two grisms (LG1 and LG2) cover the wavelength range of the channel.
Order-sorting filters were also developed to limit the grism dispersers to their desired wavelength ranges by 
rejecting higher-order shorter-wavelength light.
For SG1 and LG1 that cover short half of the MIR-S and MIR-L wavelength range, respectively, the filters block longer-wavelength lights, whereas filters for SG2 and LG2 covering the longer half block shorter-wavelength lights.
In addition, since the MIR-L and MIR-S channels are also sensitive to the MIR-S and NIR wavelength ranges, respectively, 
the SG1 and LG1 filters should also block shorter-wavelength light.
Therefore, band-pass filters were adopted for filters of both SG1 and LG1.
Note that the longer wavelength light for the SG2 is blocked by the transmission of the ZnS filter.
For the LG2 filter, the longer wavelength light is beyond the long wavelength cutoff of the Si:As detector array and the CdTe filter.

\begin{figure*}
	\begin{center}
	\FigureFile(160mm,50mm){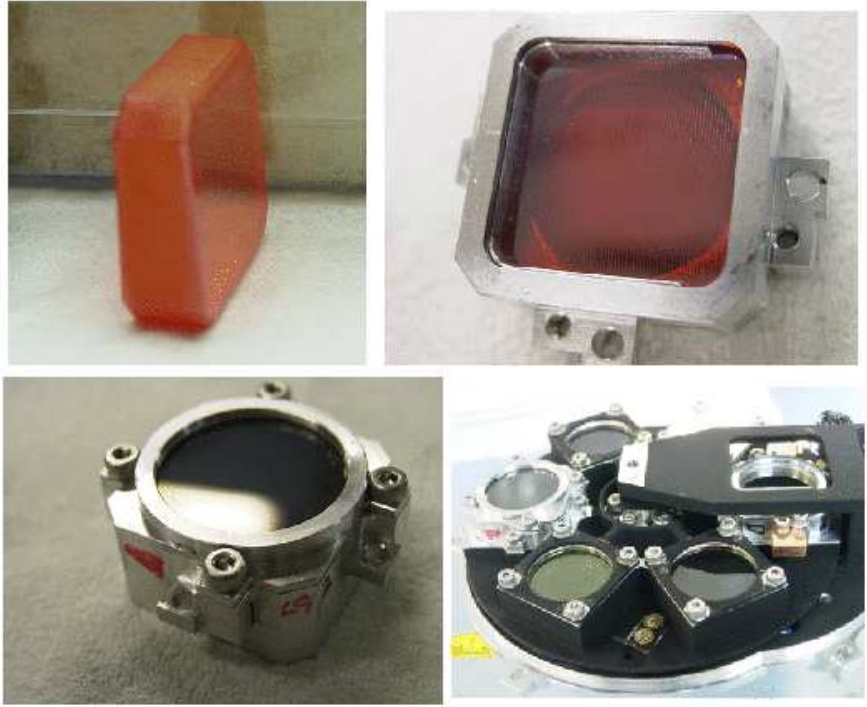}
	\end{center}
\caption{Pictures of the MIR-L grisms.
($upper$ $left$) The LG2 grism.
Dimensions of the LG2 grism are 20.0 mm, 20.0 mm, and 6.0 mm in height, width, and thickness, respectively.
($upper$ $right$) The LG2 grism stored in its holder.
($lower$ $left$) The same holder as shown in $upper$ $right$, but is seen upside-down, showing the order-sorting filter.
Dimensions of the order-sorting filter are 19.9 mm and 2.0 mm in diameter and thickness, respectively.
($lower$ $right$) The MIR-L filter wheel mounting the LG1 and LG2 as well as three imaging filters (L15, L18W, and L24).
The filter wheel has a diameter of 90.0 mm.
Note that the picture shows the filter wheel set up for laboratory tests.
MIR-S grisms (SG1 and SG2) were assembled and installed onto their corresponding filter wheels in a very similar way, 
and NG was also mounted in a comparable fashion but its order-sorting filter is directly coated on its surface.
}\label{fig:LG1}
\end{figure*}

For higher-dispersion spectroscopy in the NIR channel, we developed a grism (NG) to cover 2.5--5.0$\mu$m.
In order to avoid order wrapping of the spectral images, the shortest wavelength range (1.8--2.5$\mu$m) of the NIR channel is blocked by the order-sorting filter that is coated on the surface of NG.
The grisms and their corresponding order-sorting filters were mounted together in grism holders for SG1, SG2, LG1, and LG2 (Figure \ref{fig:LG1}), while a grism coated by its order-sorting filter was mounted in its holder for NG.
After assembling optical elements in their holders to form individual dispersing elements, they were mounted on their corresponding filter wheels.
Since these dispersers are heavier than other imaging filters, two dispersers were located at opposite sides of the wheel for better balance of their weight.
Table \ref{tab:disperser_spec} and \ref{tab:disperser_design} summarize the specifications and design of the dispersers.

\begin{table*}
\begin{center}
\caption{Specifications of dispersers.
All the values were measured in-orbit.\footnotemark[$*$]}\label{tab:disperser_spec}
\begin{tabular}{llllll}
\hline
channel      & name & effective coverage & dispersion   & $R$\footnotemark[$\dagger$] & $L$\footnotemark[$\ddagger$]  \\
            &      & $\mu$m--$\mu$m & $\mu$m/pixel &                 & pixel            \\
\hline
NIR   & NP  & 1.8--5.5  & 0.06\footnotemark[$\S$] & 19 at 3.5$\mu$m & 54 \\
            & NG  & 2.5--5.0  & 0.0097 & 120 at 3.6$\mu$m & 258 \\
MIR-S & SG1 & 4.6--9.2  & 0.057  & 53 at 6.6$\mu$m & 62 \\
            & SG2 & 7.2-13.4 & 0.097  & 50 at 10.6$\mu$m & 60 \\
MIR-L & LG1\footnotemark[$*$] & (11--19) & (0.173) & (34 at 14.4$\mu$m) & (46) \\
            & LG2 & 17.5-26.5 & 0.17  & 48 at 20.2$\mu$m & 53 \\
\hline
\multicolumn{6}{@{}l@{}}{\hbox to 0pt{\parbox{135mm}{\footnotesize
\footnotemark[$*$] LG1 is not available in orbit, and numbers shown in parenthesis are based on its design.\\
\footnotemark[$\dagger$] Spectral resolving power ($R$) in the slit-less spectroscopy mode, that is defined by $\lambda$/$\delta \lambda$, where $\delta \lambda$ is the smallest resolving element size corresponding to the FWHM of the point spread function in N3, S9W, and L18W for NIR, MIR-S, and MIR-L channels, respectively.\\
\footnotemark[$\ddagger$] Size of dispersed spectroscopy images within ranges where spectral response curves are larger than 5\% (for NP and NG) or 20\% (for SG1, SG2, and LG2) of their peaks.\\
\footnotemark[$\S$] Dispersion of NP is not linear, and the quoted value is a representative one measured at 3.5$\mu$m.
}\hss}}
\end{tabular}
\end{center}
\end{table*}

\begin{table*}
\begin{center}
\caption{Design of dispersers}\label{tab:disperser_design}
\begin{tabular}{cccccccc}
\hline
name & \multicolumn{4}{l}{dispersers} & \multicolumn{3}{l}{order-sorting filter} \\
\hline
                      & type  & material & grooves   & prism/blaze angle & material & type\footnotemark[$\dagger$] & spectral coverage \\
                      &       &          & mm$^{-1}$ & deg               &          &      & $\mu$m \\
\hline
NP                    & prism & CaF$_{2}$/Si & NA    & 56/10.7           & NA   & NA & NA      \\
NG                    & grism & Ge      & 47.6       & 2.86              & Ge   & SC & $>2.5$     \\
SG1                   & grism & KRS-5   & 10.02      & 2.6               & ZnS  & BP & 5.0--8.3 \\
SG2                   & grism & KRS-5   & 5.71       & 2.2               & ZnS  & SC & $>7.0$     \\
LG1\footnotemark[$*$] & grism & KRS-5   & 3.83       & 2.3               & CdTe & BP & 11.0--19.0 \\
LG2                   & grism & KRS-5   & 3.70       & 3.4               & CdTe & SC & $>$17.7    \\
\hline
\multicolumn{8}{@{}l@{}}{\hbox to 0pt{\parbox{105mm}{\footnotesize
\footnotemark[$*$] not available in orbit.\\
\footnotemark[$\dagger$] SC=short-end cut, BP=band-pass cut.
}\hss}}
\end{tabular}
\end{center}
\end{table*}

As for the LG1, although it worked well in the IRC's single component tests in the laboratory, it became opaque
\footnote{
The LG1 because opaque as a result of micro-cracks propagating in the sample most likely due to an unexpected abrupt change in the temperature of the cryostat during the ground test operation.
}
during {\it AKARI}'s system integration tests after assembling the entire observing system (including the telescope and two science instruments) within the cryostat, and has not been available since then.

\subsection{Aperture Masks}

In addition to slit-less spectroscopy, slit spectroscopy and higher-dispersion spectroscopy in a small aperture are also available for diffuse and point-like targets, respectively.
The focal-plane aperture masks of the IRC have unique shapes with various types of substructure at the sides of the main FOV for imaging (see \citet{onaka07} for the illustration).
Here we describe the design and purposes of the aperture masks in more detail.

There are three narrow slits for spectroscopy of diffuse sources.
The first one, the Ns slit, is a narrow slit for the NIR and MIR-S channels, while the Ls slit is another narrow slit for the MIR-L channel.
The Ns slit is shared by both NIR and MIR-S channels by means of the beam splitter.
Widths of these slits are chosen to match typical imaging PSF sizes for the channel (5\arcsec~ width for the Ns slit and 7\arcsec~ for 
the Ls slit, corresponding to the PSFs of the S9W and L18W filters, respectively) to maximize the resolving power of the dispersers.
Another slit for higher dispersion spectroscopy, the Nh slit, is created at the outer part of the Ns slit that is not covered by the 
MIR-S channel.
Its width is narrower than that of the Ns slit to match the imaging PSF size of the N3 filter (3\arcsec), giving the highest spectral resolution with the NG.
Note that the absolute pointing accuracy of the telescope is only as good as $\sim 3$\arcsec~ (\cite{onaka07}), which is not accurate enough to locate a compact source at these narrow slits and, hence, these slits are useful only for observing diffuse sources extended well over $\sim 3$\arcsec.
Since we take a direct image of the targeted area as well as spectroscopy images, without changing telescope pointing attitude (see section \ref{sec:AOT} below), the observer can obtain the slit position information from the image with an accuracy of $\lesssim 3$ arcsec.

In addition to narrow slits, another small aperture, the Np aperture, is created for spectroscopy of single targeted sources with the NG.
The smaller aperture ensures that contamination of the spectra by source overlap is minimal even for higher-dispersion spectroscopy with the NG whose spectral image length is as large as the full array size.
We do not have a 'peak-up' capability to locate an object on the slit and we totally rely on the satellite pointing accuracy.
Therefore, the size of the aperture is set as 1\arcmin~ by 1\arcmin~ according to the absolute pointing accuracy expected prior to launch taking account of possible thermal deformation of the satellite/telescope structures
\footnote{After tuning the satellite attitude control system in orbit and the ground supporting software for observation planning, the current absolute pointing accuracy of the telescope is 3\arcsec~ or better (\cite{onaka07}).}
.

\section{The Astronomical Observing Template (AOT) for Spectroscopy}\label{sec:AOT}

An astronomical observing template (AOT) for spectroscopic mode (AOT04) is designed to cover the entire spectral range of the channel within a single pointing opportunity.
To cover the entire spectral coverage of the IRC, however, one needs to make at least two pointings for a target; one for the NIR 
and MIR-S channels, and a second for the MIR-L channel.
In the first half of the pointing period, spectroscopic exposures with the first disperser (NP or NG for the NIR channel, SG1 for the MIR-S channel, and LG2 for the MIR-L channel) will be made, 
and more exposures with another disperser (NP or NG for the NIR channel, SG2 for the MIR-S channel, and LG2 for the MIR-L channel) will follow in the last half, after taking a single imaging exposure for source position reference (the ``reference image").
The sequence is schematically illustrated in \citet{onaka07}.
For slit-less spectroscopy, the reference images provide wavelength reference positions in the spectral images.
The reference images are taken with wider-band filters (S9W for the MIR-S channel and L18W for the MIR-L channel), or near the center of the wavelength coverage of the channel (N3 for the NIR channel).
The telescope pointing will not be dithered because the reference image should be taken at exactly the same position 
where spectroscopy is done.

The AOT accepts two types of options.
Since two NIR dispersers (NP and NG) are complementary to each other, i.e., both cover a similar wavelength range with different dispersions, observers need to choose one of the two according to their scientific goals.
Also, observers must specify the pointing offset preset of the telescope boresight according to their desired observing mode.
For slit-less spectroscopy, one selects offset for either the center of the NIR/MIR-S imaging FOV or the center of the 
MIR-L imaging FOV. For slit spectroscopy of diffuse sources, one selects offset for either Ns, Nh, or Ls slit.
For point-source spectroscopy with the NG, one selects the Np aperture.

\section{Performance and Calibration of the IRC Spectroscopy Mode}

\subsection{Repeatability of Disperser Insertion Position}

Repeatability of the filter-wheel positions when inserting the dispersers is very important to ensure good wavelength calibration.
Each wheel is connected directly to a stepping motor with 24 latching positions. Any error in filter wheel rotation relative
to its nominal insertion position would result in a tilt of the spectrum, leading to an erroneous wavelength calibration. 
During one year of operation of the filter wheel rotation, with over $\sim 500$ pointing opportunities for spectroscopy, 
the errors of the disperser insertion angles (or the differences in orientation of a dispersed spectroscopy image from 
its nominal orientation) have never exceeded an angle $\lesssim 1$ deg, or $\lesssim 1$ pixel shift along spatial direction over dispersed spectroscopy image length of 50--60 pixels (see Table \ref{tab:disperser_spec}).
This assures sufficient accuracy for wavelength calibration.

\subsection{Sky Brightness and Noise Conditions in Spectroscopy Images}

The slit-less spectroscopic images of the background (sky) emission (mostly zodiacal emission) are found to be as bright as that in the direct images, and are within range of our pre-launch expectation.
With the high background in the MIR, we can achieve background-photon-noise-limited conditions in the long exposure frames of $\sim 16$ seconds (see \cite{onaka07} for short- and long-exposure frames).
However, the zodiacal light is much fainter in the NIR channel, so that exposures need to be longer than 44 seconds to be background limited.
The exposure time of individual sub-frames in the spectroscopic mode is rather short; the same as that for imaging observations.
But, compared with conventional slit spectroscopic observations, several sub-frame images can be stacked with less degradation of performance by array-readout noise.

For slit spectroscopy with NP, NG, SG1, SG2, and NG spectroscopy in a small aperture (Np), 
array-readout noise usually dominates the background noise, due to the faintness of the sky per pixel.
Noise in the spectrum is dominated by background-photon noise at longer wavelengths (LG2).

\subsection{Wavelength Calibration and Its Accuracy}\label{sec:wavelength_calib}

The wavelength calibration curves (wavelength as a function of pixel) were basically measured by observing bright emission-line objects.
The results of computer ray-tracing simulation of the NIR optics were also taken into account for NP.
Most of the calibration observations were made during the initial performance verification (PV) phase, and additional observations were made
during calibration time.
One should measure calibration curves at various positions across the imaging FOV and slit areas to define the positional dependence of the calibration curves.
However, such a detailed calibration program was not possible because of the limited visibility of the calibration objects and the 
small number of pointing opportunities available for calibration during both the PV phase and calibration time.
Therefore, we measured the wavelength calibration curves either in the slits or near the imaging FOV center, and the results were checked by another measurement made 
at the other location.
An extended planetary nebula (PN) was selected for wavelength calibration in the slits, and compact sources such as Wolf-Rayet (WR) stars, compact star-forming regions, and an AGN were selected for observation near the imaging FOV center.
Due to the low spectral resolution (except for NG), one can detect only bright emission-lines with large equivalent widths even for bright objects, and the number of detectable emission lines is relatively small (2--4) (see section \ref{sec:example} and Figure \ref{fig:N6543_SPEC} for the slit spectrum of the PN).
Therefore, we adopted simple linear wavelength calibration curves for all the grisms.
The wavelength calibration curve for NP is not linear and low-resolution spectroscopy would not provide an adequate
shape for the function at shorter wavelengths, where resolution is even smaller, so we defined the shape of the calibration curve by
computer ray-tracing simulation of the NIR camera optics and the NP prism.

For NG, SG1, and SG2, a bright PN located within {\it AKARI}'s constant viewing zone (CVZ), NGC 6543, was observed through slits, and the wavelength calibration curves measured there were checked by slit-less spectroscopy of the bright WR stars, WR 128 and WR 133 (\cite{WR_cat}).
For LG2, a compact and bright AGN (Mrk 3: e.g., \cite{weedman05}) was observed in the slit-less spectroscopy mode, and the measured calibration curve was checked both by additional 
slit-less spectroscopy of bright knots (presumably compact star-forming regions) along the spiral arms of a nearby galaxy, M81,
and by slit spectra of the PN, although the latter shows only a single emission line within the range of LG2.
For NP, slit spectra of the PN were taken to check the simulated wavelength calibration curve, and a small pixel 
offset between the simulated function and the observation was determined by using a few emission lines at longer wavelengths 
where the calibration curve is closer to linear and the spectral resolution is larger.
The measured wavelength calibration curve was checked by slit-less spectra of the WR stars.
In all cases, the fitting error (residual) of the wavelength calibration curves is typically 0.5 pixel for all dispersers.

Some wavelength calibration errors are unique to our slit-less spectroscopy.
One type of error arises from uncertainties in the wavelength reference positions, since there are no such well-defined references
in IRC slit-less spectroscopy like the slits for conventional slit spectroscopy.
Although we take reference images to define wavelength reference positions of the detected sources by measuring their locations on the images, 
their positions would be shifted if the telescope pointing attitude were to drift in the wavelength direction during the pointing period
\footnote{Typical drift rate is 2\arcsec~per minute or less.
}.
As described below, the drift (or shift of image positions) could be measured by using NP or NG spectra and comparing their locations on 
the reference image. The typical accuracy of shift measurements is 0.5 pixel or less for NP and about 0.5 pixel for NG (see section \ref{sec:extraction} for more details)
\footnote{Because wavelength calibration accuracy is determined by the accuracy of correcting the drift of the telescope attitude, or the objects' movement on detector arrays, here we present the wavelength error in units of pixels.
See Table \ref{tab:disperser_spec} to find errors in $\mu$m.
}.
Another type of error occurs when comparing wavelength calibrations of objects observed at different positions across the imaging FOV and slit areas.
Since it is difficult to measure the detailed positional dependence of the calibration in orbit, we rely on pre-launch laboratory testing 
with a special pin-hole mask installed instead of the aperture mask in flight.
We took a pair of imaging and spectroscopic images of a continuum source through the grid, exactly as for the spectroscopy AOT operation in orbit, and measured positions of the holes in imaging and spectroscopy modes.
For the grisms, we measured positions of zeroth-order light images as positional references on the spectroscopic images.
By comparing their positions, we found that the pixel shift from the ideal uniform wavelength calibration is typically 0.5 pixel for all grisms.
For the prism (NP), we found that the pixel offset changes significantly over the entire array, although we could not quantitatively characterize the shift because no good positional indicator (like zeroth-order images of grism spectral images) is available in the prism spectra.
Fortunately, in orbit, we found that many blue field stars show prominent pseudo-spectral features around $2.4\mu$m (see section \ref{sec:extraction} below for more details), and we could measure the pixel shift over the entire FOV, as we did for the grisms on the grid images at the laboratory.
We successfully fitted the pixel offset by using a higher-order function of the pixel locations, with an accuracy of $\lesssim 0.5$ pixel over the entire FOV.
We confirmed that the measured function reproduces the laboratory images, though less precisely.
The pixel offset function thus measured was taken into account to represent the wavelength calibration curves and to calibrate observed spectra of NP.
We suspect that mis-alignment between the two prisms comprising the NP causes the positional dependence of the wavelength calibration.
Combining all these sources of error in fitting and creating the calibration curves, the overall accuracy of the wavelength calibration 
is typically about 1 pixel for all dispersers.

\subsection{Flux Calibration and Its Accuracy}

Flux calibration was made based on slit-less spectroscopic observations of flux standard stars, and spectral response curves were calculated based on their spectral templates 
in flux density (mJy) units.
A list of standard stars was taken from the infrared standard star networks (\cite{cohen99}; \cite{cohen03}; \cite{cohen07}), and we selected our targets from the list by considering visibility and brightness of the stars (table \ref{tab:stdstars}).
Several stars are common to the IRAC flux calibration program onboard {\it Spitzer} (\cite{reach05}).
In order to check the validity of the spectral templates, we observed at least two stars per disperser; one an $A$-type dwarf showing only 
Hydrogen absorption lines, and the other a $K$-type giant showing several broad absorption features from 2--10$\mu$m, to validate the calibration for each disperser, except for the LG2 where only one $K$-type star is available.
Since we found no significant systematic differences in the spectral response curves among the observations (above the 10\% level), spectral response curves that were measured with brighter stars ($K$-type stars for NP and LG2, and $A$-type stars for other grisms) were adopted.

\begin{table*}
\begin{center}
\caption{Standard stars for flux calibration}\label{tab:stdstars}
\begin{tabular}{lllll}
\hline
star & 2MASS J & type & $A$v (mag) & target dispersers \\
\hline
2MASS J17564393+6920089 &                  & A4 V     & 0.508 & NP \\
KF06T2                  & 17583798+6646522 & K1.5 III & 0.189 & NP \\
KF09T1                  & 17592304+6602561 & K0 III   & 0.000 & NG/NP \\
Bp+66 1073              & 18030959+6628119 & K1 III   & 0.000 & NG/SG1/SG2 \\
NPM 1+67.0536           & 17585466+6747368 & K2 III   & 0.161 & NG/SG1/SG2 \\
HD42525                 & 06060937-6602227 & A0 V     & 0.000 & NG/SG1/SG2/LG2 \\
HD42701                 & 06065055-6716599 & K3 III   & 0.604 & LG2 \\
\hline
\end{tabular}
\end{center}
\end{table*}

\begin{figure}
	\begin{center}
	\FigureFile(80mm,50mm){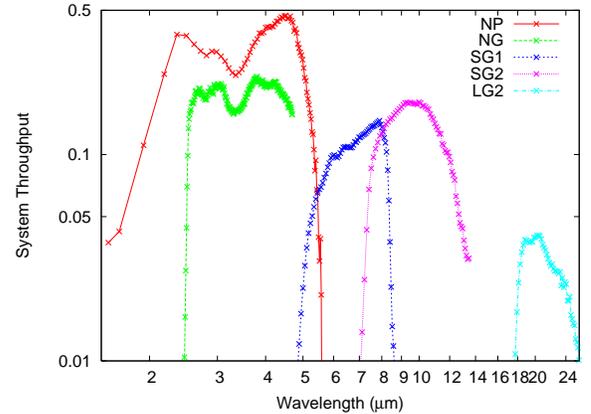}
	\end{center}
\caption{System throughput of the IRC spectroscopy mode.
}\label{fig:throughput}
\end{figure}

\begin{figure}
	\begin{center}
	\FigureFile(80mm,50mm){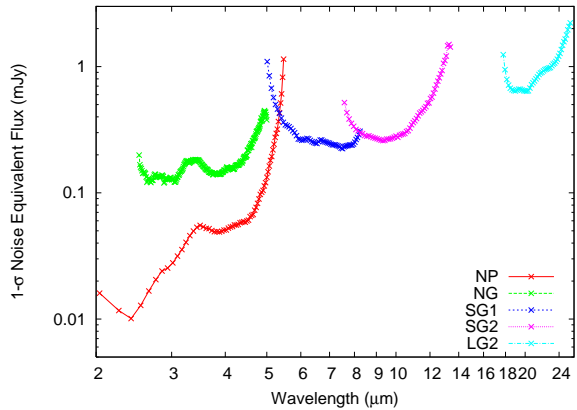}
	\end{center}
\caption{Sensitivity as a function of wavelength.
The plot shows one-sigma noise-equivalent flux (in mJy) at the ecliptic poles as a function of wavelength.
Because of smaller dispersion at shorter wavelength in NP, the NP has higher sensitivity at shorter wavelength.
}\label{fig:noiseplot}
\end{figure}

The spectral response curves were then used to derive the end-to-end system throughput of the {\it AKARI}/IRC in orbit, including telescope optics, camera optics of the IRC with disperser inserted, and quantum efficiency of detectors (Figure \ref{fig:throughput}).
The system throughput is well within the range of pre-launch expectation based on throughput measurements of all the individual optical 
components as well as the quantum efficiency of the detectors.
A dip in the throughput around 3.4$\mu$m found in both NP and NG originates from the beam splitter.
The sensitivity is calculated based on the system throughput and the observed noise levels in orbit.
Here, the noise is a typical fluctuation of the sky counts on the stacked and fully-calibrated images, after masking out all the source spectra, and the measurement was done after smoothing over $2\times 2$ pixels for all dispersers to roughly match their spectral and spatial resolutions.
The one-sigma noise-equivalent flux spectrum for slit-less spectroscopy at the ecliptic poles, where background zodiacal light is faintest, is shown in Figure \ref{fig:noiseplot}.
The sensitivity agrees well with the range of the pre-launch expectation based on the expected system throughput as well as readout- and sky-photon noises for all grisms except for NP.
For NP we found that the sensitivity is worse than expected by a factor of 2 or more, although the system throughput at the spectral range is as good as expected.
This is because the spectra of many faint field stars can be detected with NP, especially around 2.4$\mu$m where the sensitivity per 
pixel is largest, amplifying the sky fluctuations around any target spectrum.

Monitoring of sensitivity stability in both imaging and spectroscopic modes has been on-going since the initial PV phase.
Because the IRC spectroscopic mode shares most of the optical components of its imaging mode (except for imaging filters and dispersers), and the operation of the detector arrays is exactly the same for both modes, results of the 
sensitivity monitoring in the imaging mode should provide us with a good measure of the sensitivity stability in the spectroscopic mode.
Results of the imaging sensitivity monitor are briefly described in \citet{onaka07}. It is stable at a level of 5\% over almost a year in all the filters.
Results of more detailed studies will be presented elsewhere (\cite{tanabe07}).
Sensitivity monitoring of the spectroscopic mode has been conducted less frequently.
Although it is difficult to perform accurate sensitivity monitoring in the slit-less spectral mode, no systematic trend in sensitivity has been 
found at the 20\% level in all the dispersers.
In summary, so far there are no indications of temporal variations of the sensitivity in either the imaging or spectroscopic modes.

There is a noteworthy influence of wavelength calibration error on the flux calibration for slit-less spectroscopy.
Because of the very low dispersion and wide spectral coverage with single dispersers, the spectral response per pixel changes 
greatly near the ends of their spectral coverage.
In addition, due to the slit-less nature of this IRC spectroscopic mode, wavelength calibration accuracy is not as good ($\sim 1$ pixel; see section \ref{sec:wavelength_calib}).
As a result, flux calibration errors that arise from the wavelength mismatch between the observed spectra and the spectral response curve could be much larger near the band ends than around the band center.
The data reduction tool kit (see below) incorporates this kind of error when plotting spectra with error bars.

After considering all sources of errors as mentioned above, the overall flux calibration accuracy is estimated to be 10\% for all dispersers around their central wavelengths, and it gets worse (20\%) around longest and shortest wavelengths of the spectral coverage of each disperser.

\section{Data Reduction}

The IRC team has been developing a software suite (the so-called toolkit) for reducing the IRC spectroscopic images.
The software is written in IDL, and utilizes The IDL Astronomy User's Library at Goddard Space Flight Center, NASA\footnote{http://idlastro.gsfc.nasa.gov/}.
With the software, one first performs a series of basic data reductions of the array images (dark subtraction, linearity correction, correction of various kinds of image anomaly, etc.) to derive basic-calibrated images, and then applies spectroscopy-specific calibration on the basic-calibrated images to derive spectroscopic products.
Since the array operation and data acquisition sequence (or the AOT operation) in the spectroscopic mode follow exactly the same pattern as that for the imaging mode, the same techniques can be used for the basic reduction of the array images in the spectral images.
Note, however, that we are still investigating the causes of, and correction methods for, anomalies, and currently the anomaly correcting routines are not fully implemented.
See \citet{onaka07} for more details on the basic data reduction.
Here we describe only the spectroscopy-specific calibrations.
First, we describe data reduction techniques that are unique to slit-less spectroscopy, and then comment on calibration techniques common to conventional slit spectroscopy.

\subsection{Extracting the spectral images for each source}\label{sec:extraction}

Finding accurate wavelength reference positions is the key to deriving well-calibrated spectra from slit-less spectroscopy, because any drift in telescope pointing would cause errors in wavelength calibration and, hence, flux calibration that requires the wavelength-dependent spectral response function.
Since source positions that are measured on the reference image provide the wavelength reference positions on the spectral images, the reference and spectral images should be taken at exactly the same telescope pointing.
In reality, however, because pointing drift is often observed within a single pointing period, the amount of drift should be measured by using the IRC data themselves and be corrected accordingly.
First, we apply a cross-correlation technique to find the pixel shift between spectral sub-frame images, and the shift is corrected before stacking them.
Next, we need to align the stacked spectral image to the reference image to find accurate wavelength reference positions as well as the locations of extraction boxes for each source.
For this purpose, we detect field stars and compare their positions in reference and spectral images, since field stars can be observed at any sky orientation in the NIR channel although they are much fainter in the MIR-S and MIR-L channels.
Along the spatial direction, it is simple to measure shifts because spectra of continuum point sources show a rather 
sharp ridge structure that can be traced in the spatial direction.
Along the wavelength direction, shifts can be measured by using the characteristics of spectral response curves that create pseudo-spectral features in data number, before flux calibration.
For NP, spectra of most field stars show prominent peaks around 2.4$\mu$m where the throughput plot shows a local peak (Figure \ref{fig:throughput}) because the NP has a lower dispersion at shorter wavelengths, and typical field stars have blue colors in the wavelength range.
For NG, since the response curve shows a rather sharp cut-off to its shortest wavelength range (around 2.5$\mu$m: Figure \ref{fig:throughput}) due to an order-sorting filter, an abrupt drop in data number spectra can be seen at that wavelength for most field stars.
These features change little in shape for most stars, regardless of their spectral types, because the Rayleigh-Jeans part of the spectrum 
is observed across the wavelength range.
For MIR-S and MIR-L channels, the pixel shift measured at the NIR channel is adopted, after considering the pixel scale differences.
In addition to these spectral features, one can also detect point-like images of zeroth-order light for brighter point sources even in the spectral 
images with the grisms, providing another method to find and correct the pointing attitude drift in the wavelength direction.

\subsection{Flat-fielding}

There are some difficulties in flat-fielding of the slit-less spectroscopic images.
First, we should apply a kind of aperture correction for spectral images of the background light that is not required for spectra of compact sources.
Note that the slit-less spectroscopic image of even the spatially-flat background is not flat at all due to the overlapping of zeroth- 
and higher-order light images with the first-order light image (for grism spectroscopy images), as well as dispersed `edge' images of the imaging aperture (for all dispersers).
Here, ``aperture correction" implies flattening the background sky image before removing the background contribution underlying the object spectra.
In addition to the aperture correction, we should treat the color-dependence of the flat response, since IRC spectroscopy covers a wide 
spectral range simultaneously, and the color-dependence of the flat response cannot be neglected.
Therefore, in order to take all these flat characteristics into account, we apply a flat-field correction twice; first for aperture correction and pixel-to-pixel flat correction of the background, and then for wavelength-dependent flat-field correction (or color correction) for individual compact objects.
For the aperture correction and flat-field correction of the background, the sky-flat images were created by stacking large numbers of observed spectral images with three-sigma clipping rejection algorithm so as to remove any overlaid compact sources as well as cosmic-ray hits.
The flattened background is then subtracted from the observed slit-less spectroscopic images.
After extracting spectral images for each source and calibrating their wavelengths, color correction can be applied throughout each extracted source spectrum.
Here, the amount of the color correction for a given pixel is estimated by using sky-flats for two different imaging filters 
(N3 and N4 sky-flats for color-correcting NP and NG; S7 and S15 sky-flats for color-correcting 
SG1 and SG2; and L15 and L24 sky-flats for color-correcting LG2), and interpolating two sky-flat images along the wavelength direction for the pixel.

\subsection{Spectroscopy calibrations common to conventional slit spectroscopy}

After extracting spectral images of each object, the two axes of the extracted images correspond to the spatial and wavelength directions.
Therefore, the following calibration can be made in a similar way to that for conventional slit spectroscopy.
The wavelength calibration can be achieved simply by applying the wavelength calibration curve to each extracted spectrum in the same way as for the slit spectroscopy.
After the color correction, spectral sensitivity is only a function of wavelength, and flux calibration can also be made in the same way as for slit spectroscopy.

Finally, we remark on data reduction for IRC slit spectroscopy.
Reducing slit spectroscopy images can be done more simply than for the case of slit-less spectroscopy, since we do not have to take care of extraction, wavelength calibration, and color-correction of individual source spectra according to the locations of the sources.
On the other hand, one must take much more care in subtracting the contribution of scattered light within a camera body and/or 
within an array substrate, especially for MIR-S and MIR-L channels, in analyzing slit spectral data.
One obvious reason is that the background level is much lower in the slit areas.
Another reason that is unique to the IRC is that a larger number of sky photons directly illuminate the detector arrays even in 
slit spectroscopy through the imaging FOV adjacent to the slit areas.
Note that the scattered light is not spatially uniform due to aperture substructures and edges around the detector arrays, making the estimation of the scattered light contribution difficult.
See \citet{sakon07} for more discussion and techniques for removing the scattered light contribution in analyzing images in slit spectroscopy.

\section{Examples of the IRC Spectra}\label{sec:example}

To demonstrate the IRC's spectroscopy capability, we present spectra of a compact source, UGC 5101, observed with the slit-less spectroscopic mode, and of an extended source, NGC 6543 (PN), observed in slits.

\begin{figure*}
	\begin{center}
	\FigureFile(105mm,150mm){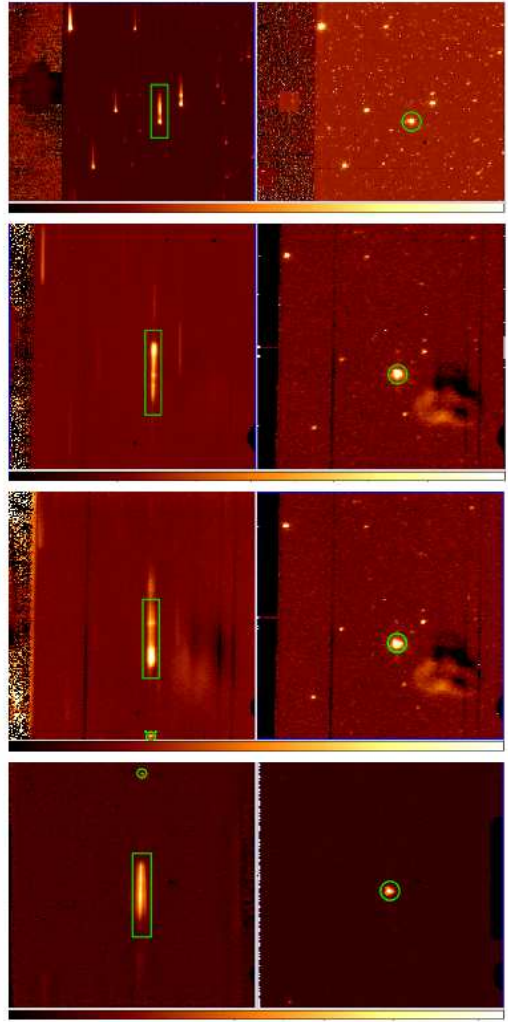}
	\end{center}
\caption{Images for the slit-less spectroscopy observations of UGC 5101.
Images shown are basic-processed data (after dark subtraction, linearity correction, flat fielding, and stacked among frames, etc.) of spectroscopy ($left$) and reference ($right$) images.
Spectral images of NP ($top$), SG1, SG2, and LG2 ($bottom$) are shown on the $left$ side, and reference images of N3 ($top$), S9W, S9W, and L18W ($bottom$) are shown on the $right$ side.
The slit areas are located on the $left$ side of NIR and MIR-S images, and on the $right$ side of MIR-L image.
The spectra and reference images of UGC 5101 (a bright object located near the center of each image) are marked in $green$ on all the images.
Wavelength increases upward in NP, SG1, and SG2, and downward in LG2.
The zeroth-order light images are seen below the object in SG2, and above the object in LG2, and they are marked in $green$ circles.
The zeroth-order light image is located outside the FOV in SG1.
The second-order light images, that are not marked, can also be seen at just above the object in SG1 and SG2, and at just below the object in LG2.
Some kinds of anomaly (dark vertical stripes and an incorrectly flattened area at lower right) are also seen, especially in S9W, SG1, and SG2 images.
Note that size of the NIR images (NP and N3) are reduced to be half of the MIR-S and MIR-L images for clarity of the figures.
}\label{fig:U5101_IMAG}
\end{figure*}

\begin{figure*}
	\begin{center}
	\FigureFile(120mm,50mm){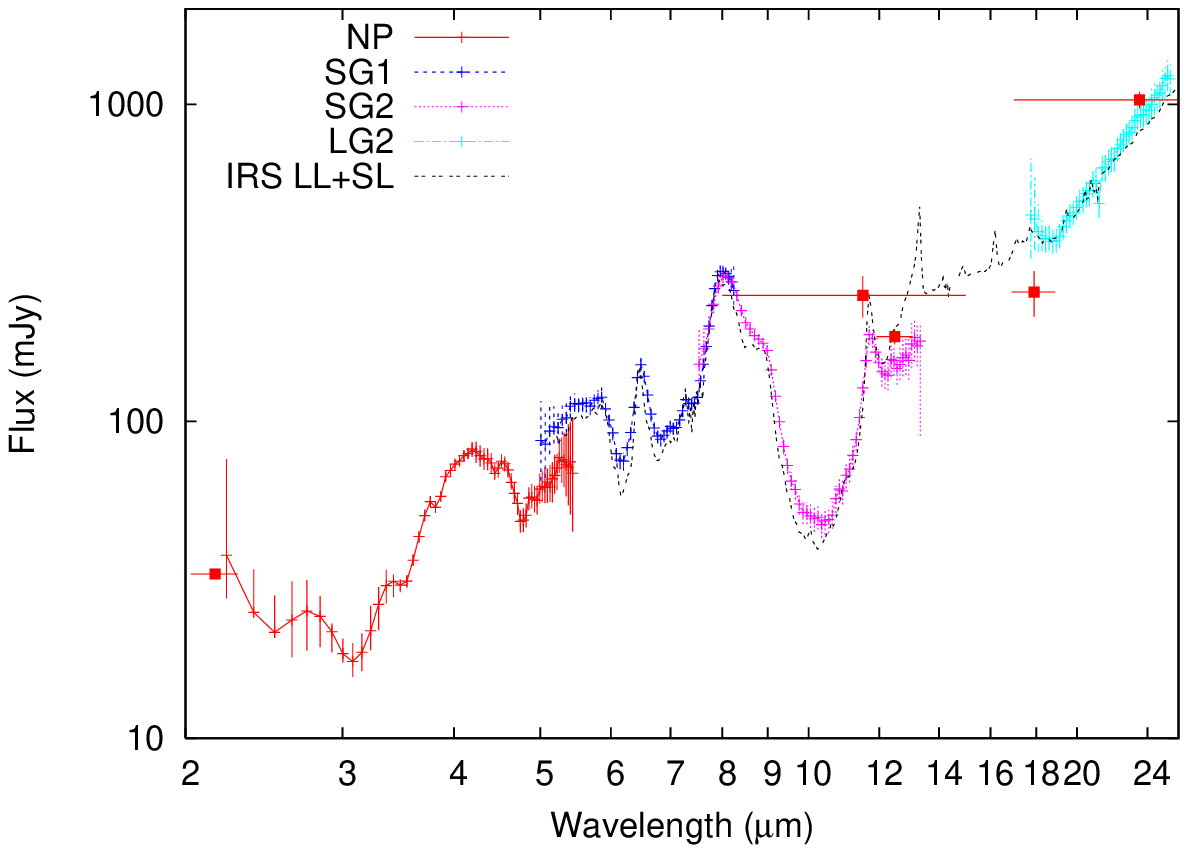}
	\end{center}
\caption{Extracted one-dimensional spectrum of UGC 5101.
The four spectra of NP, SG1, SG2, and LG2 are shown in $red$, $blue$, $magenta$, and $cyan$, respectively.
The one-sigma error bars are also shown.
Some photometry data taken from literature are overlaid in $red$, that are 2MASS ($K_{\rm s}$), ground-based 12.5$\mu$m and 17.8$\mu$m, and IRAS 12 and 25$\mu$m data.
The IRS spectrum, taken with low-resolution short- and long-wavelength modules, is also overlaid in $black$ broken lines on the plot.
}\label{fig:U5101_SPEC}
\end{figure*}

UGC 5101 is a heavily obscured ultra-luminous infrared galaxy (e.g., \cite{genzel98}) at a redshift of $z=0.0394$.
Recent MIR spectroscopic study of the galaxy, made with the IRS onboard {\it Spitzer} (\cite{irs}), revealed that it shows rich absorption features as well as PAH emission bands (\cite{armus04}), both of which create a series of broad bumps across the NIR and MIR spectral ranges.
Because of these prominent spectroscopic characteristics, we choose this object to check the wavelength and flux calibrations of the IRC spectroscopic mode that were measured separately as described in the previous sections.
The observation was made on April 22, 2006, during the PV phase.
Two pointed observations were performed, once with a target at Nc and another at Lc, to cover the entire spectral range of the IRC.
We choose AOT04 operation with the NP in the NIR channel for both these observations.
Figure \ref{fig:U5101_IMAG} shows the entire spectral and reference images, before extracting the target's spectrum, to illustrate basic ideas 
on the appearance of the images in slit-less spectroscopy.
Since the object is very bright in the MIR, zeroth- and second-order light images can be also seen (except for the SG1 image where the zeroth-order light is located outside the FOV).
Figure \ref{fig:U5101_SPEC} shows the extracted spectrum.
The spectrum was compared with previous photometric results available in the literature (2MASS (\cite{2mass_catalog}) data in the NIR, 
IRAS (\cite{iras_es}) and ground-based 12.5$\mu$m and 17.8$\mu$m MIR data (\cite{soifer00}), and we found that the IRC spectrum is basically consistent with this photometry.
The IRS spectrum (\cite{irs}), the ``post-BCD'' spectra (program id: 105; AOR KEY: 4973056) obtained from the {\it Spitzer} science archive
\footnote{The NASA/IPAC Infrared Science Archive is operated by the Jet Propulsion Laboratory, California Institute of Technology, under contract with the National Aeronautics and Space Administration.}, is overlaid for comparison.
We found that the two spectra agree very well, indicating that both wavelength and flux of the IRC spectra are well calibrated 
when compared with the precursor.
Detailed discussion of the detected spectral features will be made elsewhere (\cite{nakagawa07}).

\begin{figure*}
	\begin{center}
	\FigureFile(120mm,50mm){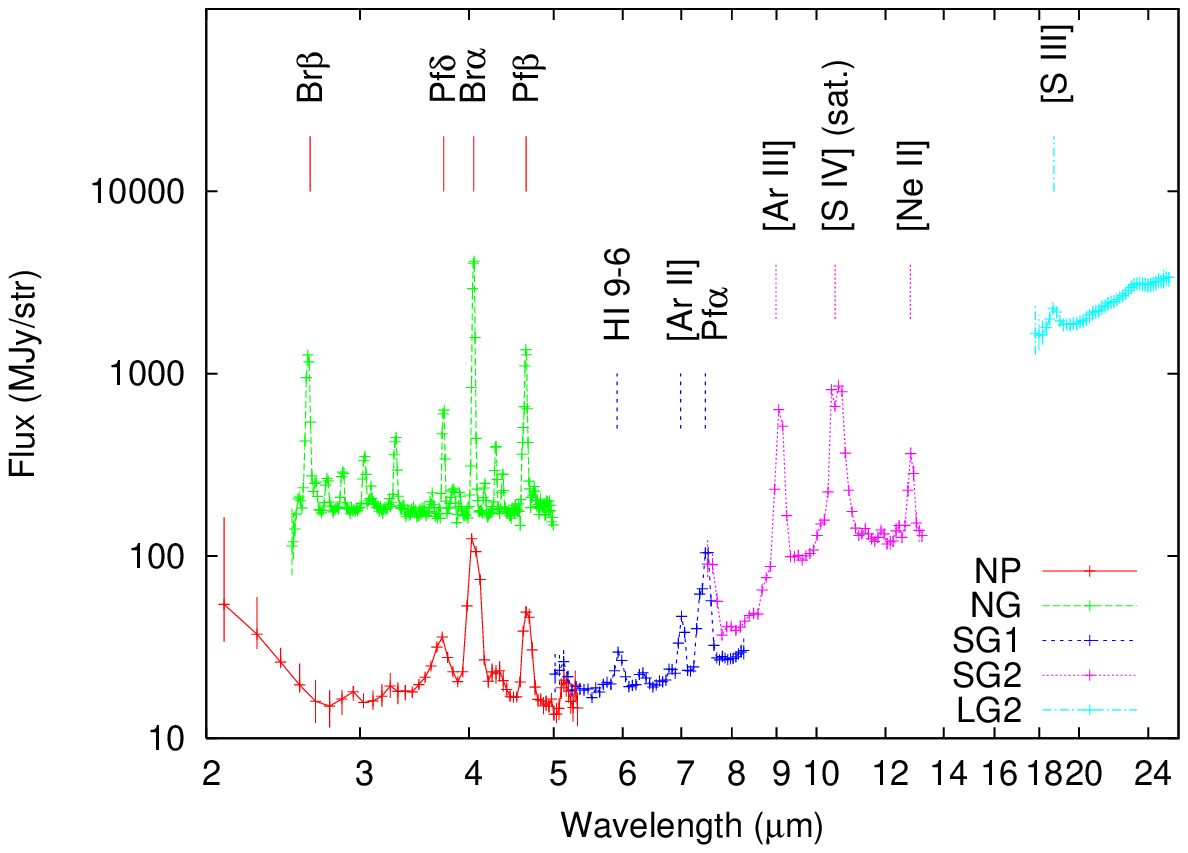}
	\end{center}
\caption{The slit spectrum of NGC 6543.
The spectra of NP, NG, SG1, and SG2 were taken at the Ns slit, while that of LG2 was taken at the Ls slit.
The NG spectrum is multiplied by 10 for clarity of the plot.
At the NIR channel, since the spectra were taken at wider Ns slit, emission lines are as wide as $\sim 4$ pixel, being wider than the spectra at the narrower Nh slit whose width is $\sim 2$ pixel.
The [S IV] emission line is saturated.}\label{fig:N6543_SPEC}
\end{figure*}

NGC 6543 is a bright PN located near the north ecliptic pole.
Because of its brightness and convenient location for observing with space telescopes in near-earth polar-orbit, this object was used for calibrating both flux and wavelength of mid-infrared instruments on IRAS (e.g., \cite{iras}), ISO (\cite{n6543_sws_wlcal}), and {\it Spitzer} (\cite{irs}), as well as for the study of its physical conditions (e.g., \cite{bernard-salas03}).
As expected, the observed spectrum (Figure \ref{fig:N6543_SPEC}) reveals rich emission lines, including Hydrogen recombination lines and fine-structure lines of various excitation levels.
We have successfully identified all the bright emission lines reported in \citet{bernard-salas03}.
We found that emission-line widths are very close to the slit widths measured in the reference images, indicating that all the dispersers work well in orbit, delivering their expected resolving powers.

\section{Summary}

The Infrared Camera (IRC) on board the {\it AKARI} satellite has the capability of spectroscopy as well as multi-color photometry.
In this paper, its spectroscopic mode is described in detail in terms of dispersers and aperture masks with slit areas, in-orbit calibration, operation, data reduction, and overall performance with examples of spectra.
In particular, we have described the characteristics of its unique slit-less spectroscopy observation mode that was made available for the first time from space in MIR.
We found that the overall in-orbit performance of the IRC spectral mode is well within expectation from pre-launch experiments, and it satisfies its original design goal except for loss of one disperser, the LG1.
So far about 500 pointed observations were made with IRC spectroscopy within a year of {\it AKARI}'s science operation without any serious troubles, and one may expect scientific results from the spectroscopic data very soon.

\bigskip

AKARI is a JAXA project with the participation of ESA.
We thank all the members of the AKARI project for their continuous help and support.
W. K., I. S., and T. T. have been financially supported by the Japan Society of Promotion of Science (JSPS).
This work is supported in part by a Grant-in-Aid for Scientific Research from the Ministry of Education, Culture, Sports, Science, and Technology of Japan, and Grants-in-Aid for Scientific Research from the JSPS (Nos. 16077201, 16204013, and 17740122).

\end{document}